\def\wide#1#2{
\end{multicols}
\vspace{-0.1in}
\widetext
\noindent
\if#1t
\else
\raisebox{9pt}[0in][0.0in]
{$\rule{3.4in}{0.4pt}\rule{0.4pt}{6pt}$\hspace{3.6in}}
\fi
\vspace{-0.15in}
#2
\vspace{-0.2in}
\if#1b
\else
\raisebox{-9pt}[0in][0.0in]
{\hspace{3.55in}$\rule{0.4pt}{6pt}\rule[6pt]{3.5in}{0.4pt}$}
\fi
\begin{multicols}{2} 
\vspace{-0.05in}
\noindent
}
\documentstyle[aps,prl,multicol,epsf]{revtex}
\begin{document}
\author{A.\ E.\ Koshelev and I.\ S.\ Aranson}
\address{Materials Science Division, Argonne National Laboratory, 
Argonne, Illinois 60439}
\title{Resonances, instabilities, and structure selection of driven Josephson
lattice \\
in layered superconductors}
\date{\today}
\tighten
\maketitle
\begin{abstract}
We investigate dynamics of Josephson vortex lattice in layered high
T$_{c}$ superconductors at high magnetic fields.  It is shown that
the average electric current depends on the lattice structure and is
resonantly enhanced when the Josephson frequency matches the frequency
of the plasma mode.  We find the stability regions of moving lattice. 
It is shown that a specific lattice structure at given velocity is
uniquely selected by the boundary conditions:
at small velocities periodic triangular lattice is stable and looses
its stability at some critical velocity.  At even higher velocities a
structure close to a rectangular lattice is restored.
\end{abstract}

\pacs{74.60.Ge, 74.50.+r,47.54.+r}
\vspace{-0.4in}
\begin{multicols}{2}

Transport properties of layered superconductors, 
such as Bi$_{2}$Sr$_{2}$CaCu$_{2}$O$_{x}$ (BSCCO),  in magnetic field
parallel  the layers are determined by the dynamics of Josephson
vortex lattice (JVL). 
\cite{LeeAPL95,HechPRL97,LatyshPhysC97} Moving JVL generates travelling
electromagnetic wave in the medium.  Similar to the Eck resonance in a
single junction\cite{Eck}, one expects a strong resonance emission when
the velocity of the lattice matches the plasma wave
velocity.\cite{KoySSC95,ArtJETPL97} In the current-voltage (I-V)
dependence this resonance is seen as a strong enhancement of current at
fixed voltage.  In contrast to a single junction, JVL in layered
superconductors has soft degrees of freedom related to the phase shifts
between different layers.  This leads to a rich variety of
dynamic states observed in numerical simulations.\cite{JVLSimul} In
particular, a periodic lattice corresponds to a constant phase shift
between neighboring layers, ranging from $0$ for the rectangular lattice
to $\pi$ for the static triangular lattice.  Moving lattice generates
electromagnetic wave with the $c$-axis wave vector selected by the
lattice structure.  Since the velocity of a plasma wave propagating along
the layers depends on this wave vector \cite{SakaiPRB94}, \emph{the
resonant velocity of the lattice depends on its structure}.
%

In this Letter we investigate the stability of moving JVLs and the evolution of
structure as a function of its velocity for large size
samples.\cite{fiske} We show that a specific lattice structure at given
velocity is uniquely selected by the boundary conditions.  At small
velocities a periodic lattice is stable.  The phase shift between
neighboring layers smoothly decreases with the increase of velocity,
starting from $\pi$ for a static lattice.  At some critical velocity the
lattice becomes unstable.  At even higher velocities a periodic lattice
with the phase shift smaller than $\pi/2$ is restored again.

Consider a layered superconductor in a magnetic field applied along the
layers ($y$ axis) with transport current flowing across the layers ($z$
axis).  The dynamics of such a system is described by the inductively coupled
sine-Gordon equations for the gauge-invariant phase differences $\theta
_{n}$.\cite{PhaseDynEq} Introducing the reduced time $\tau =\omega
_{p}t$ and in-plane coordinate $u=x/(\gamma s)$, we express currents
$j_{x,n}$, $j_{z,n}$ and electric fields $E_{z,n}$ via $\theta _{n}$ and
the reduced magnetic fields $h_{n}=H_{n}2\pi \gamma \lambda_{ab}
^{2}/\Phi _{0}$ as
\begin{eqnarray}
&&j_{z,n} =j_{J}\left (\sin \theta _{n}+ \nu_{ab}\partial \theta _{n}/\partial 
 \tau \right), \; E_{z,n}\approx E_{p}  \partial \theta _{n}/\partial \tau,\nonumber\\
&&j_{x,n+1}- j_{x,n}=j_{ab}\left (1+\nu_{ab}\frac{\partial}{\partial 
 \tau}\right)\left( \frac{\partial
 \theta _{n}}{\partial u}-\frac{h_{n}}{l^{2}}\right),\nonumber 
\end{eqnarray}
where $\omega_{p}$ is the Josephson plasma frequency, $\gamma$ is the
anisotropy factor, $s$ is the interlayer spacing, $j_{J}$ is the
Josephson current, $E_{p}\equiv \Phi _{0}\omega _{p}/(2\pi cs)$,
$l=\lambda _{ab}/s$, and $j_{ab}\equiv c \Phi_{0}/(8\pi
\lambda_{ab}^{2}\gamma s)$.  $\nu _{c}=4\pi \sigma _{c}/(\varepsilon
_{c}\omega _{p})$ and $\nu _{ab}=4\pi \sigma _{ab}\lambda
_{ab}^{2}\omega _{p}/c^{2}$ are the dissipation parameters, which are
determined by the quasiparticle conductivities $\sigma_{c}$ and
$\sigma_{ab}$ and are connected as $\nu _{c}/\nu _{ab}=\sigma _{c}\gamma
^{2}/\sigma _{ab}$.  Using the above relations we derive from the
Maxwell equations the coupled equations for $\theta _{n}$ and $h_{n}$
\begin{eqnarray}
\frac{\partial ^{2}\theta _{n}}{\partial \tau ^{2}}+\nu
_{c}\frac{\partial \theta _{n}}{\partial \tau }+\sin \theta _{n}
-\frac{\partial h_{n}}{\partial u} &=&0,  \label{RedTheta} \\
\left( \nabla _{n}^{2}-\frac{1}{l^{2}}\right)
h_{n}+\frac{\partial \theta _{n}}{\partial u}
+\nu_{ab}\frac{\partial }{\partial \tau } \left( \frac{\partial
\theta _{n}}{\partial u}-\frac{h_{n}}{l^{2}}\right) &=&0.
\label{RedH}
\end{eqnarray}
where $\nabla _{n}^{2}h_{n}\! \equiv\! h_{n+1}\! +\! h_{n-1}\! -\! 2h_{n}$.  Equivalent
forms of these equations have been derived in Refs.\
\onlinecite{BulPRB96,ArtJETPL97}.  In the case of negligible in-plane
dissipation ($\nu _{ab}\! =\! 0$) they can be reduced to equations containing
only $\theta _{n}$.\cite{PhaseDynEq} Taking parameters typical for underdoped
BSCCO at $T\! \approx \! 50$K: $\gamma \! =\! 500$, $\lambda _{ab}\! =\! 240$ nm, $s\! =\! 15$
\AA , $\sigma _{ab}\! =\! 2\cdot 10^{4}$ $[\Omega \cdot {\rm cm}]^{-1}$, and
$\sigma _{c}\! =\! 2\cdot 10^{-3}$ $[\Omega \cdot {\rm cm}]^{-1}$, one obtains
estimates: $\nu _{ab}\! \approx \! 0.1$ and $\nu _{c}\! \approx \! 0.002$, which we
will use in numerical computations.  For these parameters the dissipation in 
a wide range of electric and magnetic fields is mainly determined by the 
in-plane channel. 

We consider the situation when all junctions are filled by JVs.  This is
always the case at high magnetic field $B$ above $\Phi _{0}/5.5\gamma
s^{2}$ \cite{DenseJosLat} or at high currents, when all junction are
driven into the resistive state.  Neglecting self-field of the transport
current, we represent a solution to Eqs.\ (\ref{RedTheta}) and
(\ref{RedH}) for the resistive state in the form
\begin{eqnarray}
\theta _{n}(\tau ,u) &=&\omega _{E}\tau +k_{H}u+\phi _{n}(\tau )
+\tilde{\theta}_{n}(\tau ,u) \label{SolTheta}\\
h_{n}(\tau ,u) &=&h+\tilde{h}_{n}(\tau ,u).
\label{SolH}
\end{eqnarray}
Here the Josephson frequency $\omega_E$ is determined by the electric
field, $\omega_E=E_{z}/E_{p}$, and the wave
vector $k_H $ is determined by the magnetic field, $k_H=2\pi H\gamma
s^{2}/\Phi _{0}=h/l^{2}$.  $\tilde{\theta}_{n}(\tau ,u)$ and
$\tilde{h}_{n}(\tau ,u)$ are the oscillating phase and magnetic field
induced by Josephson coupling,
$\tilde{\theta}_{n},\tilde{h}_{n}\propto \sin (\omega _{E}\tau +k_{H}u
+\alpha_{n})$.  Solution (\ref{SolTheta}) corresponds to the lattice
moving with velocity $\omega _{E}/k_{H}$.  The structure of the lattice is
determined by the phase shifts $\phi _{n}$.  Without  Josephson
coupling  the system is degenerate with respect to arbitrary phase
shifts $\phi _{n}$.  Josephson coupling eliminates this degeneracy. 
It either leads to slow dynamics of phase shifts $\phi _{n}$ or selects
a certain  steady state structure.  The  equations for
$\phi _{n}(\tau)$ can be obtained using an expansion with respect to
Josephson coupling and averaging over fast degrees of freedom.  This
procedure will be described in detail elsewhere.\cite{JosLatLong}

\begin{figure}
\epsfxsize=3.2in \epsffile{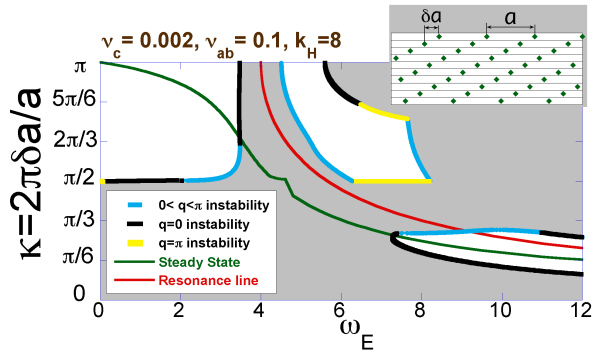}
\caption{Stability regions of moving Josephson lattice in the plane
$\omega_{E}$-$\kappa$ calculated for representative parameters
$\nu_{c}=0.002$, $\nu_{ab}=0.1$, and $k_{H}=8$.
Grey regions correspond to unstable lattices.  Sections of the boundaries
marked by black correspond to the long-wave instability.  Sections
marked by light grey correspond to the instability with $q=\pi$.  Line
starting at ($0$, $\pi$) shows dependence of the lattice wave vector on
the frequency $\omega_{E}$ selected by the boundary with free space.  We
also show the resonance line, corresponding to matching between the
Josephson frequency and the frequency of the plasma wave at the wave
vector $\kappa$.  Inset sketches a steady state corresponding to a
regular lattice.
}
\label{Fig-GenPhaseDiag}
\end{figure}

For periodic JVL one has $\phi _{n}=\kappa n$.  In such a state the
chains of JVs in the neighboring layers are shifted by the fraction
$\kappa /2\pi $ of the lattice constant (see inset in Fig.~\ref
{Fig-GenPhaseDiag}).  In particular, $\kappa =0$ corresponds to a
rectangular lattice, and $\kappa =\pi $ corresponds to a triangular
lattice (ground state at $\omega _{E}=0$).  In first order with respect
to Josephson current, the oscillating phase $\tilde{\theta}_{n}$ is
given by $\tilde{\theta}_{n}={\mathrm Im}\left [{\cal G}(\kappa )\exp
\left ( i(\omega _{E}\tau +k_{H}u+\kappa n) \right ) \right ] $ with
\[
{\cal G}(\kappa) 
\! =\! \left( \omega_E
^{2}-i\nu _{c}\omega_E -\frac{k_H^{2}\left( 1+i\nu _{ab}\omega_E
\right) }{2(1\! -\! \cos \kappa)+
\left( 1\! +\! i\nu _{ab}\omega_E \right)\! /l^{2}}\right) ^{\! -1}\!.
\]
To second order we obtain the average reduced Josephson current
$i_{J}\equiv i_{J}(\kappa ,k_{H},\omega _{E})=\left\langle \sin \theta
_{n}(\tau ,u)\right\rangle $.\cite{BulPRB96} In the case of not very
large in-plane dissipation $\nu _{ab}\omega _{E}\ll 2l^{2}(1-\cos \kappa
)+1$, it has the resonant dependence on the Josephson frequency $\omega
_{E}$ (electric field)
\[
i_{J} =\frac{1}{2}\frac{\omega _{E}\nu (\kappa )}{\left( \omega
_{E}^{2}-\omega _{p}^{2}(\kappa )\right) ^{2}+\left( \omega _{E}\nu
(\kappa )\right) ^{2}},
\]
where
$
\omega _{p}(\kappa )=k_{H}\left (2\left( 1-\cos \kappa \right)
+\frac{1}{l^{2}}\right )^{-1/2}
$
is the plasma frequency at the wave vector $\kappa $ and
\[
\nu (\kappa )\equiv \nu _{c}+\frac{2\left( 1-\cos \kappa \right)
k_{H}^{2}\nu _{ab}}{\left( 2\left( 1-\cos \kappa \right) +1/l^{2}
\right) ^{2}}
\]
is the dissipation parameter of the plasma mode.\cite{BulPRB96} When the
frequency $\omega _{E}$ matches the corresponding plasma wave frequency
$\omega _{p}(\kappa )$, a resonance enhancement of the current is
expected.  The moving lattice generates a travelling electromagnetic
wave with the Poynting vector ${\bf P}$.  For the in-plane component of
${\bf P}$ we obtain
\begin{eqnarray}
P_{x}\approx  \frac{P_{ab}\omega _{E} \omega _{p}^{2}(\kappa)/k_{H}}
{\left( \omega _{E}^{2}-\omega _{p}^{2}(\kappa )\right)
^{2}+\left( \omega _{E}\nu (\kappa )\right) ^{2}},\; 
P_{ab}=\!\frac{\Phi _{0}^{2}\omega _{p}}{32\pi ^{3}\lambda ^{2}s\gamma }.
\label{Poyntingx}
\end{eqnarray}
For typical BSCCO parameters  at low temperatures ($\lambda \! =\! 200$ nm,
$\gamma \! =\! 500$, $\omega _{p}/2\pi \! =\! 150$ GHz) we obtain 
$P_{ab}\! \approx \!
125$W/cm$^{2}$.

To investigate the stability of JVL we perturb the lattice solution as
$\phi _{n}(\tau )=\kappa n+{\rm Re}[u_{q}\exp (\alpha (q)\tau +iqn)] $
and derive from Eqs.\ (\ref{RedTheta},\ref{RedH}) the eigenvalue $\alpha
(q)\equiv \alpha(q,\kappa,k_{H},\omega_E)$
\begin{equation}
\alpha (q)=-\frac{1}{4}\frac{{\cal G}(\kappa +q)+{\cal G}^{\ast }(\kappa
-q)-2\mathop{\rm Re}\left[ {\cal G}(\kappa )\right] }{\nu _{c}+\frac{1}{4i}
\left[ {\cal F}\left( \kappa +q,\kappa \right) -{\cal F}^{\ast }\left(
-\kappa +q,\kappa \right) \right] }  \label{alphaq}
\end{equation}
with
\wide{m}{
\[
{\cal F}(q,q_{1}) \equiv \left( -2\omega _{E}+i\nu _{c}+\frac{
ik_{H}^{2}\nu _{ab}}{2(1-\cos q)+\frac{1+i\nu _{ab}\omega _{E}}{l^{2}}}\frac{
2(1-\cos q_{1})}{2(1-\cos q_{1})+\frac{1+i\nu _{ab}\omega _{E}}{l^{2}}}
\right) {\cal G}(q){\cal G}(q_{1}).
\]}
The lattice is stable if there is no exponentially growing solution in the
whole $q$-interval, i.e., $\mathop{\rm Re}[\alpha (q)]\leq 0$, for
$0\leq q \leq \pi$.  The onset of instability is characterized by the wave vector
$q_{i}$ of the most unstable mode.  There are three special cases: the
long-wave instability $q_{i}=0$, the short-wave instability $q_{i}=\pi$,
and the instability with $0<q_{i}<\pi$.  One can also distinguish
absolute and convective instabilities.\cite{conabs} 
The spectrum $\alpha (q)$ has an important symmetry property
$\alpha(q=\pi,\kappa=\pi/2)=0$.  This means that a stable region cannot 
cross the line $\kappa=\pi/2$ and \emph{it is impossible to evolve
continuously from the static triangular lattice to the fastly moving
rectangular lattice without intersecting an instability boundary}.  The
triangular lattice always looses its stability before reaching the
resonance $\omega _E = k_{H}/2$, at the frequency $\omega _{\Delta
}(k_{H})$ which is determined by a simple analytical equation \newline
$\omega _{\Delta }^{2}\!  -\!  \frac{k_{H}^{2}}{4}=\!  -\omega _{\Delta
}\left( \nu _{c}+ \frac{k_{H}^{2}\nu _{ab}}{4}\right) \!  \left(
\sqrt{1+\nu _{ab}^{2}\omega _{\Delta }^{2}}-\nu _{ab}\omega _{\Delta
}\right).$

To explore the stability of JVL we calculated numerically $\mathop{\rm Re}
[\alpha (q)]$ (\ref{alphaq}) throughout the $\omega _{E}-\kappa $ plane and
found the lines at which either $\mathop{\rm Re} [\alpha (q)]=0$ at
finite $q$ or $\mathop{\rm Re} [d^{2}\alpha (q)/dq^{2}]=0$ at $q=0$. 
The stability diagram for representative parameters $\nu
_{c}=0.002$, $\nu _{ab}=0.1$, and $k_{H}=8$ is shown in Fig.\
\ref{Fig-GenPhaseDiag}.  For these parameters we found three stability
regions at moderate $\omega _{E}$: (i)the low-velocity region
located below the resonance line and at $\pi /2<\kappa <\pi $, (ii)the
high-velocity region located along the resonance line with $\kappa $
approaching $0$ with increase of $\omega _{E}$, and (iii)the region
located above the resonance line and at $\pi /2<\kappa <\pi $ (this
region disappears at higher fields).  At the boundary of the first
region the lattice experiences a long-wave instability for $\kappa
>2.04$.  At smaller $\kappa $ the instability occurs at finite wave
vector $q=q_{i}$ and $q_{i}$ grows continuously with decrease of 
$\kappa$.

Selection by the system of a specific wave number $\kappa$ from the
continuous spectrum is not directly related to stability of the
corresponding periodic structure.  For the static case the structure is
selected by the minimum energy condition.  Such a condition is absent in
the dynamic case.  In this case a specific JVL structure can be
determined by the boundary.  To demonstrate this, we consider a
semi-infinite stack of junctions with $n=1,2\ldots $ separated by a
sharp boundary from the medium with arbitrary electromagnetic
properties.  To derive equations for the phase shifts $\phi _{n}$ for
such system we have to find a solution of the linear equations without
Josephson coupling taking into account the boundary conditions.  For plasma
wave with given frequency $\omega =\omega_{E}$ and wave vector along the
layers $k=k_{H}$ the oscillating phases ($\tilde{\theta}_{n}$) and
magnetic fields ($\tilde{h}_{n}$) in the junctions can be written as
\begin{equation}
\tilde{\theta}_{n},\tilde{h}_{n}\propto \exp (-iq_{+}n)+{\cal B}\exp
(iq_{+}n),\text{ at }n\geq 1  \label{hjunc}
\end{equation}
where $q_{+}\equiv q_{+}(k,\omega )$ is (complex) wave vector
given by
\begin{equation}
\cos q_{+} =1-\frac{k^{2} \left( 1+i\nu _{ab}\omega \right) }
{ 2\left(\omega ^{2}-i\nu _{c}\omega \right) }
+\frac{1+i\nu _{ab}\omega }{2l^{2}},
\label{qplus}
\end{equation}
with $\mathop{\rm Im} [q_{+}]>0$. The properties of the boundary
are completely characterized by the complex amplitude of reflected
wave ${\cal B}\equiv {\cal B}(k,\omega )$, which has to be found
by matching the solution (\ref {hjunc}) with electromagnetic
oscillations in the medium at $z<0$.  In general, ${\cal B}
(k,\omega )$ can be a complex number with arbitrary absolute value. Only in
the simplest case of vanishing dissipation and propagating wave ($
\mathop{\rm Im}(q_{+})=0$), ${\cal B}(k,\omega )$ 
determines a conventional reflection coefficient, $R(k,\omega )=\left| {\cal 
B}(k,\omega )\right| ^{2}$, and has property $\left| {\cal
B}(k,\omega )\right| <1$. A large class of boundaries, including
boundary with free space, is well described by the ideal
reflection ${\cal B}=-1$.  Averaging with respect to oscillating
phases and field we derive the following equation for the steady
state phase shifts $\phi _{n}$
\begin{equation}
\frac{1}{2}\sum_{m=1}^{\infty }\mathop{\rm Im}\left[ G(n,m)\exp \left(
-i\left( \phi _{n}-\phi _{m}\right) \right) \right] =i_{J},
\label{StStateBound}
\end{equation}
where
\begin{eqnarray}
G(n,m)&=&G_{0}(n-m)+{\cal B}G_{0}(n+m),
\label{RespSurfGen}\\
G_{0}(n)&\equiv& G_{0}(n;k_{H},\omega_{E} ) =
\label{RespFun_n} \\
&&\frac{\delta _{n}}{\omega _{E}^{2}-i\nu _{c}\omega _{E}}-
\frac{k_{H}^{2}\left( 1+i\nu _{ab}\omega _{E}\right) }{\left( \omega
_{E}^{2}-i\nu _{c}\omega _{E}\right) ^{2}}\frac{\exp iq_{+}|n|}{2i\sin
q_{+}}.
\nonumber
\end{eqnarray}
The second term in Eq.\ (\ref{RespSurfGen}) describes the surface
contribution and vanishes at $n,m\rightarrow \infty $.  The solution,
corresponding to the lattice in the bulk, has the form $\phi
_{n}=\kappa n+u_{n}$, where $u_{n}$ is the surface deformation,
$u_{n}\rightarrow 0$ at $n\rightarrow \infty $.
Substituting this ansatz into Eq.\ (\ref{StStateBound}) we obtain a
nonlinear degenerate system of equations for $u_{n}$.  The solution for
$u_{n}$ exists only for special values of $\kappa $, i.e., {\em the bulk
structure is selected by the boundary conditions}.  For such selected 
states $z$ component of the Poynting vector $P_{z}$ is always directed from the 
boundary towards the bulk of the sample. \cite{OppositeCase} A similar 
pattern selection mechanism  is relevant 
for various nonequilibrium  systems.\cite{DynStructSel}
At small $\omega _{E}$ the system becomes linear with respect to $u_{n}$
and its solvability condition can be found analytically.  In the case
${\mathcal B}=-1$ we obtain deformation of the lattice at small
velocities $\kappa \approx \pi -\left( \nu _{ab}+7\nu
_{c}/k_{H}^{2}\right) \omega _{E} $.  In finite system with identical
boundaries the configuration is typically symmetric because each
boundary selects the same wave number.  The waves collide in the middle
forming a phase defect.  \cite{DynStructSel}

To find the steady state configurations at all velocities we solved
Eq.$\,$(\ref {StStateBound}) numerically taking ${\mathcal B}\!  =\! 
-1$ and using the same parameters ($\nu _{c}\!  =\!  0.002$, $\nu
_{ab}\!  =\!  0.1$).  The dependence $\kappa (\omega _{E})$ obtained
from these solutions for $k_{H}\!  =\!  8$ is shown in
Fig.$\,$\ref{Fig-GenPhaseDiag} together with the stability regions.  At
small velocities the lattice experiences smooth evolution of structure
with the lattice wave vector $\kappa $ decreasing from $\pi $ at zero
velocity to smaller values until it hits the instability boundary.  At
higher velocities the stable lattice with $\kappa \!  <\!  \pi /2$ is
restored.  The structure continues to evolve smoothly towards the
rectangular configurations with increase of velocity.  Near the line
$\kappa \!  =\!  \pi/2$ we observe transition to the double-periodic
lattice $\phi_{n}\!  =\!  \pi n/2 \!  +\!  (-1)^{n}v$, which becomes
stable at high fields.

\begin{figure}
\epsfxsize=3.4in \epsffile{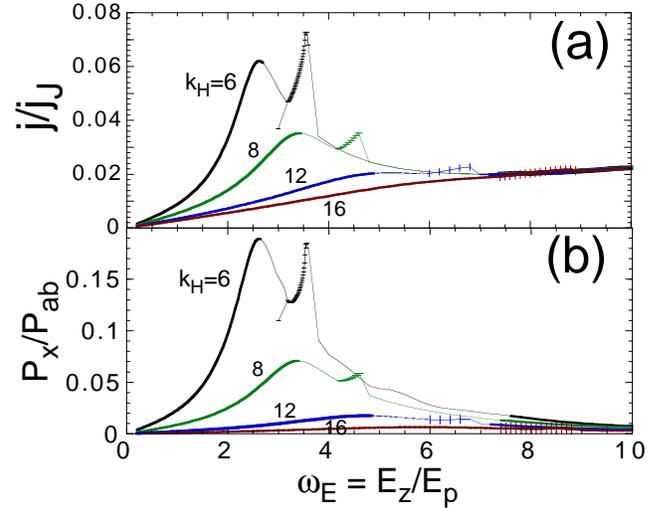} \caption{(a) Current-voltage
characteristics at different magnetic fields, for $k_{H}=6$, $8$, $12$,
and $16$ (for $\gamma= 500$ $H\approx k_{H}\cdot 0.3$T).  The
dependencies are obtained using numerically computed steady states with
parameters $\nu_{c}=0.002$, $\nu_{ab}=0.1$.  Thick lines show stable
branches and thin lines show unstable branches.  The branches, marked by
dashes, correspond to the double-periodic lattices.  (b) Electric field
dependencies of the Poynting vector along the layers for electromagnetic
wave generated by the moving lattice (see Eq.\ (\protect
\ref{Poyntingx})).  The scale $P_{ab}$ is defined in Eq.\ (\protect
\ref{Poyntingx}).}
\label{Fig-IV}
\end{figure}

\begin{figure}
\epsfxsize=3.2in \epsffile{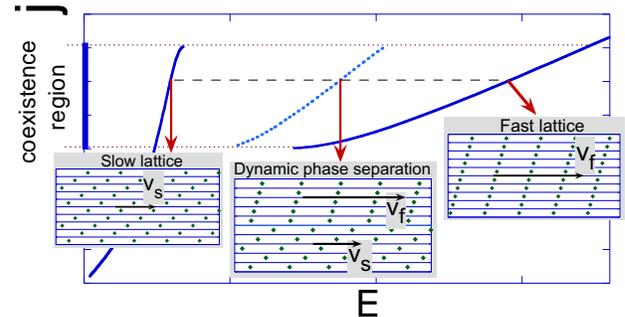} 
\caption{Multibranch structure of the current-voltage characteristic due
to dynamic phase separation.  Two states corresponding to slow lattice
motion (velocity $v_{s}$) and fast lattice motion (velocity $v_{f}$)
coexist within the current range marked at the vertical axis.  In
this region the intermediate phase-separated states exist, in which the
system is split into fastly and slowly moving regions.  The intermediate
branch corresponding to one of such states is shown by dotted line.}
\label{Fig-PhaseSep}
\end{figure}

Fig.\ \ref{Fig-IV} shows evolution of the current-voltage dependence and
the in-plane Poynting vector (\ref{Poyntingx}) with increase of magnetic
field.  Up to magnetic field
$H\approx \sqrt{\sigma_{ab}/(\gamma^2\sigma_{c})}\Phi_0/(\pi\gamma 
s^2)$ ($k_{H}\approx 16$), the
current-voltage dependencies have two stable branches, corresponding to
moving periodic lattices, separated by broad instability region where
periodic JVL can not exist.  In this regime two lattice solutions exist
within a finite range of currents.  Such coexistence is facilitated by
the high in-plane dissipation, which causes strong dependence of the JVL
velocity on its structure.  Within the coexistence region one can expect
a family of intermediate states, in which the system is split into two
(or more) domains moving with different velocities separated by a phase
defect (dynamic phase separation, see Fig.\ (\ref {Fig-PhaseSep})).  The
phase-separated states give the most natural interpretation of the
multiple I-V branches observed by Hechtfischer {\it et al.}, who studied
transport properties of JVL in BSCCO mesas at high magnetic
fields.\cite{HechPRL97} This interpretation can be verified by measuring
the spectrum of microwave irradiation emitting by the stack.  Instead of
a single peak located at the Josephson frequency corresponding to the
average voltage, the spectrum of irradiation should contain two peaks
corresponding to the ``fast'' and ``slow'' states.

In conlusion, we investigated stability and boundary structure selection
of the driven JVL. We found two major stability regions, separated by
unstable region: the low-velocity region correspondsto moving
structure close to triangular lattice and the high-velocity region
corresponds to an almost rectangular lattice.

AEK thanks R.\ Kleiner, N.\ F.\ Pedersen, M.\ Tachiki, and K.Gray for helpful
discussions.  This work was supported by the U.S. DOE, Office of Science
under contract \# W-31-109-ENG-38.  AEK also would like to acknowledge
support from the JST (Japan) and to thank the National Research Institute
for Metals for hospitality.  
\vspace{-.2in}
\vspace{-.1in}
\vspace{-.3in}
\end{multicols}

\begin{references}
\bibitem{OppositeCase}
The opposite case describes the scattering of the incoming wave generated
far away from the boundary with the wave vector given by an external source
(or opposite boundary).  In this case the boundary does not select the
structure in the bulk.  \protect \cite{DynStructSel}
\end{references}
\end{document}